# A GPU-based Multi-level Algorithm for Boundary Value Problems

J. T. Becerra-Sagredo, F. Mandujano and C. Málaga

October 5, 2016

**Abstract**

A novel and scalable geometric multi-level algorithm is presented for the numerical solution of elliptic partial differential equations, specially designed to run with high occupancy of streaming processors inside Graphics Processing Units (GPUs). The algorithm consists of iterative, superposed operations on a single grid, and it is composed of two simple full-grid routines: a restriction and a coarsened interpolation-relaxation. The restriction is used to collect sources using recursive coarsened averages, and the interpolation-relaxation simultaneously applies coarsened finite-difference operators and interpolations. The routines are scheduled in a saw-like refining cycle. Convergence to machine precision is achieved repeating the full cycle using accumulated residuals and successively collecting the solution. Its total number of operations scale linearly with the number of nodes. It provides an attractive fast solver for Boundary Value Problems (BVPs), specially for simulations running entirely in the GPU. Applications shown in this work include the deformation of two-dimensional grids, the computation of three-dimensional streamlines for a singular trifoil-knot vortex and the calculation of three-dimensional electric potentials in heterogeneous dielectric media.

Parallel multigrid; GPU; CUDA

## 1 Introduction

The multi-grid algorithm [4, 27] is one of the fastest methods for solving linear and non-linear systems of equations derived from a variety of problems, like numerical discretizations of partial differential equations and non-linear variational problems [18, 21]. The main idea is to accelerate the convergence of an iterative method using a hierarchy of nested grids where discretizations perform resolution-dependent corrections that are passed between levels using interpolations. Multi-grid has the main advantage over other methods that when used to solve the problems with a given accuracy, its number of operations often scale linearly with the number of discrete nodes used. It has been widely studied and optimized for sequential operation, providing very fast convergent algorithms for a wide variety of applications [6, 22, 28, 5, 23, 19].

Multi-grid algorithms can be roughly divided in two groups: geometric multi-grid (GMG) and algebraic multi-grid (AMG). GMG works over a regular grid which can be perfectly subdivided into lower resolution grids, whereas AMG works over unstructured grids without a fixed neighbor node structure. GMG methods are more efficient than AMG on structured problems, since they take advantage of the fixed geometric representation, which is likewise thought to limit their applicability. However, they can handle some complex geometries by implementing boundary conditions over immersed interfaces [1].

Early works on multi-grid for GPUs date back to 2003, when GPUs started to outperform CPUs and control over the operations and memories of the GPU were available using programmable vertex shaders [3, 12]. These implementations are many-core maps of a classic multi-grid algorithm using recursive V-cycles. Additional interesting works in this direction are found in [26, 16, 13, 25], and more recently using Nvidia's compute unified device architecture (CUDA) in [20, 8, 7, 10, 24, 17].

A consequence of the use of a classic multi-grid in many-core architectures is the appearance of performance penalties for coarse grids, where the number of independent operations is reduced and memory latencies cannot be hided using multithreading. In order to avoid these penalties, a hybrid approach is preferred by many authors, see [11, 9], where the CPU is used for serial matrix inversions over coarse grids, and the GPU for computing fine-grid relaxations. This is a natural way to avoid the low parallelism of the coarse grids but with the penalty of the memory exchange between the random access memory (RAM) of the CPU and that of the GPU, which can be several orders of magnitude lower in bandwidth than the GPU fast off-chip memory.

In this work we present a novel geometric multi-level algorithm that performs fully parallel operations over single grid data without the need of multiple grids. The algorithm is designed to work entirely inside the GPU without data transfers to the CPU. It is implemented with fine-grain parallelism, using a thread per grid node, such that its performance scales with every new generation of many-core architectures. It has only two simple full-grid many-core kernel routines: the relaxation-interpolation and the restriction. The routines are scheduled in a refining cycle, from the coarsest to the finest scales. Convergence of the solution of the linear system to machine precision is achieved repeating the cycle using the accumulated residual and collecting the overall result. A novel radial finite difference stencil is proven to provide convergence when used for the discretization of the heterogeneous Helmholtz-Poisson equation.

The algorithm is presented in section 2 and results are presented in section 3. The results include convergence tests, and some applications to two-dimensional grid deformation, three-dimensional streamlines of a singular trifoil-knot vortex and three-dimensional electric potentials in heterogeneous dielectric media.

## 2 The Algorithm

Let the elliptic partial differential equation be denoted by

$$\mathscr{L}(u) = f, \tag{1}$$

where $\mathscr{L}$ is a spatial partial differential operator in a bounded, cubic domain $\Omega \in \mathbb{R}^2$ or $\mathbb{R}^3$, with Dirichlet, Neumann or mixed boundary conditions for $u$ at the boundary $\partial\Omega$.

The solution of (1) is equivalent to the steady state solution of the pseudo-time dependent problem

$$\frac{\partial u}{\partial \tau} = (\mathscr{L}(u) - f), \tag{2}$$

with respective boundary conditions, starting, for example, from null fields.

The time explicit numerical integration of (2) requires a large amount of steps to reach a steady state due to the slow transfer of low frequency components of the solution [4]. Therefore, analogous to classic multi-grid, we propose to solve first (2) for the low frequency components over a coarse subset of the entire mesh and interpolate the result to the rest, repeating the process using the advanced solution data in finer subsets. For this, as in the case of GMG, we need a grid that allows nesting, with $N = 2^n + 1$ points per dimension, and $n \in \mathbb{N}$. The nested subset of level $v$ has $2^{n-v} + 1$ grid points and spacing of $2^v h$, such that $v = 0, 1, ..., n-1$.

But instead of using residuals and errors to collect the solution over an extended set of multiple nested grids, like in a classic multi-grid algorithm, we have chosen to superpose operations over a single grid data, fully in parallel. Obtaining the solution and interpolating it to the rest of the mesh, almost simultaneously. Notice that this strategy only makes sense for a fully parallel implementation and provides a completely different method to a classic multi-grid. What we call the single-grid multi-level algorithm (SGML) contains only two full-grid many-core kernel routines: the restriction and the relaxation-interpolation, organized in a refining cycle with recurrences. This strategy allows for the occupation all the streaming processors of the GPU, without the penalties associated to low number of threads on coarse grids, and without communications to the CPU.

The restriction is a local volume average of given fields, such as sources and coefficients, necessary to correctly solve the problem for a given level $v$. The idea is to represent the fields in a coarse subset of the mesh by widening the support of interpolations given by tensor products of linear interpolations, known as hat functions $\wedge(x)$.

```
1:    g₁ = f
2:    do m from 0 to ν
3:        λ = 2^m
4:        for every (i, j, k)
5:            g₂ = ℛ_λ(g₁)
6:        g₁ = g₂
```

Algorithm 1. The restriction($\nu$).

The sum of the neighboring sources, evaluated using the widened hat functions, can be done directly or using a simple multi-dimensional geometric reduction, consisting of a sequence of nested sums, from the finest grid till the desired level $\nu$, as shown in Algorithm 1. In line 5 of Algorithm 1, $\mathscr{R}_\lambda(f)$ is the weighted average of neighbors at distance $\lambda h$. In two dimensions

$$\mathscr{R}_\lambda(f_{i,j}) = \frac{1}{4} \sum_{p=-1}^{1} \sum_{q=-1}^{1} \wedge\left(\frac{-p}{2}\right) \wedge\left(\frac{-q}{2}\right) f_{i+\lambda p, j+\lambda q}, \quad (3)$$

where $\wedge(\pm 0.5) = 0.5$, and in three dimensions

$$\mathscr{R}_\lambda(f_{i,j,k}) = \frac{1}{8} \sum_{p=-1}^{1} \sum_{q=-1}^{1} \sum_{r=-1}^{1} \wedge\left(\frac{-p}{2}\right) \wedge\left(\frac{-q}{2}\right) \wedge\left(\frac{-r}{2}\right) f_{i+\lambda p, j+\lambda q, k+\lambda r}. \quad (4)$$

In the SGML, we choose the application of (2) over a coarse subset of the grid, with an explicit discretization using maximum stable pseudo time step and radial finite differences for $\mathscr{L}(u)$, denoted by $\delta\mathscr{L}(u)$. The relaxation-interpolation performs two operations in parallel for different sets of grid points as shown in Algorithm 2. The relaxation subset grid is defined by a multi-dimensional power of two module (line 3 of Algorithm 2), and if a node belongs to this subset, then the relaxation operation $\mathscr{D}_\lambda$ is applied (line 4 of Algorithm 2). $\mathscr{D}_\lambda$ represents the discretization of equation (2) in pseudo-time, with restricted sources and coefficients, using a forward Euler advance with stable time step and finite differences approximation of the spatial operator $\mathscr{L}$ with grid spacing $\lambda h$. The stability condition for $\mathscr{L} = \nabla^2$, with second order finite differences $\delta\mathscr{L}(u)$, is a pseudo-time step $\delta\tau < (2^\nu h)^2/2$.

The interpolation operator $\mathscr{I}_\lambda$ (line 6 of Algorithm 2) acts over the complement of the relaxation subset grid and uses the tensor product of hat functions $\wedge(x)$ with support $2\lambda h$ to interpolate the variation $\delta u = \mathscr{D}_\lambda(u_1) - u_1$, obtained for those nodes in the relaxation subset grid at the previous iteration.

The exploration of refining cycles has provided us with the fast convergence cycle described in Algorithm 3 and sketched in Figure 1. It's a saw-like pattern that is resemblant, but not equivalent, to what is known as a full multi-grid cycle [4]. The number

```
1:   λ = 2^ν
2:   u_1 = u
3:   if (i, j, k) = (0, 0, 0) mod λ
4:      u = 𝒟_λ(u_1)
5:   else
6:      u = 𝓘_λ(δu_1)
```

Algorithm 2. The relaxation-interpolation($\nu$).

```
1:   n = log_2(N − 1)
2:   do ν_1 from n-1 to 0
3:      do ν from ν_1 to ν_1 − 1
4:         restriction(ν) [Algorithm 1]
5:         do min(n_r, 2^{n−ν_1}) times relaxation-interpolation(ν) [Algorithm 2]
6:      do ν = ν_1
7:         restriction(ν) [Algorithm 1]
8:         do min(n_r, 2^{n−ν_1}) times relaxation-interpolation(ν) [Algorithm 2]
9:      do ν = 0
10:        do min(n_r, 2^n) times relaxation-interpolation(0)
```

Algorithm 3. The single cycle.

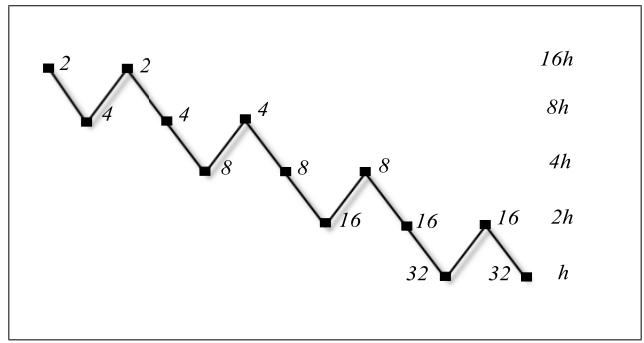

Figure 1: Single cycle with doubling relaxation-interpolations for a $33^2$ mesh. The numbers beside the points indicate the number of relaxation-interpolations per level for full doublings. During the recurrence of cycles, this number can be bounded by a maximum number of relaxations $n_r$.

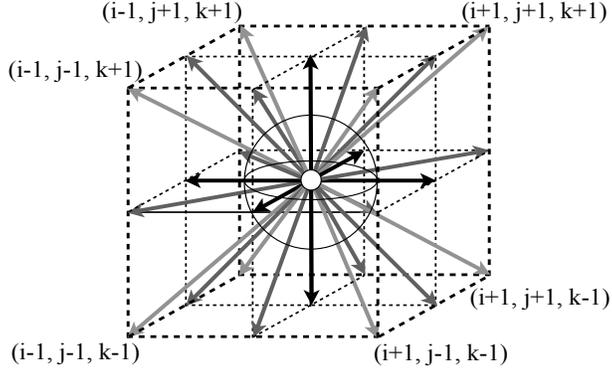

Figure 2: Scheme for the radial finite differences approximation of the divergence of the gradient (5) at coordinates $(i, j, k)$. Individual radial derivatives are computed with all the 26 neighbors and integrated over the area of the sphere with radius $h/2$ using the divergence theorem.

of relaxation-interpolations doubles for every level $v$ and can be limited by a maximum number of repetitions $n_r$ (see lines 5, 8 and 10 of Algorithm 3).

The convergence achieved in a single cycle is extended to a desired tolerance by recurring the cycles using the residual after each cycle as source term and adjusting the boundary conditions. The cycles are repeated, first substituting the sources with the residual $r_1 = f - \delta \mathscr{L}(u_0)$, and so on successively, solving each recurrence for $e_i$ in $\delta \mathscr{L}(e_i) = r_i$, where the residual of cycle $i$ is $r_i = r_{i-1} - \delta \mathscr{L}(e_{i-1})$. If we define $r_0 = f$ and $u_0 = e_0$, then $u = \sum_{i \geq 0} e_i$ is the solution of the algebraic system to machine precision for enough iterations.

Dirichlet boundary conditions are imposed in the variation $\delta u$ for the first cycle. Neumann boundary conditions require the use of images in the restriction routine, the imposition of the derivative at the boundary substituting data lying outside of the domain, and the integral of the sources over the whole domain to be zero. The latest condition is necessary to avoid the migration of the solution, given that the Neumann problem has multiple shifted solutions.

## 3 Numerical examples

Consider the Helmholtz equation in heterogeneous media with $\mathscr{L} = \nabla \cdot (\sigma \nabla) + a$. Convergence of the algorithm has been achieved using a discretization with a 9-point stencil in two dimensions or a 27-point stencil in three dimensions. The 27-point formula,

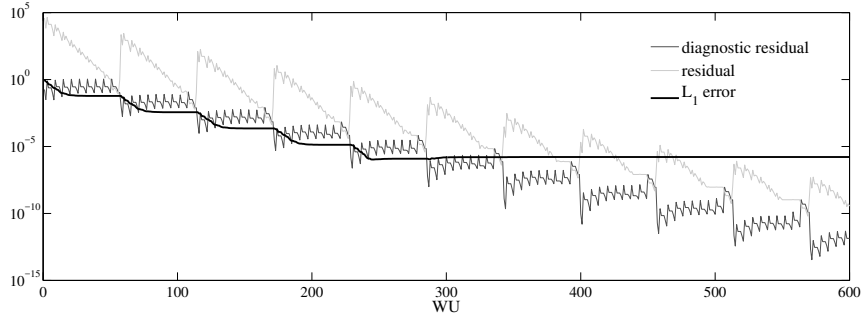

Figure 3: Convergence of the SGML algorithm for the Poisson equation with source (8). The mesh is $1025^2$. The cycles perform only two relaxation-interpolations per level. Full mesh operations are counted as working units.

sketched in Figure 2, is given by

$$\nabla \cdot (\sigma \nabla u)_{i,j,k} \approx \frac{3}{13h} \sum_{p=-1}^{1} \sum_{q=-1}^{1} \sum_{r=-1}^{1} \frac{(\sigma_{i+p,j+q,k+r} + \sigma_{i,j,k})}{2} \frac{(u_{i+p,j+q,k+r} - u_{i,j,k})}{hl_{p,q,r}}, \quad (5)$$

where $l_{p,q,r}$ is unity for $(\pm 1, 0, 0)$, $(0, \pm 1, 0)$, $(0, 0, \pm 1)$; $\sqrt{2}$ for all combinations of $(\pm 1, \pm 1, 0)$, $(\pm 1, 0, \pm 1)$, $(0, \pm 1, \pm 1)$; and $\sqrt{3}$ for all combinations of $(\pm 1, \pm 1, \pm 1)$.

## 3.1 Convergence

The SGML is tested for convergence, solving the two-dimensional Poisson equation

$$\nabla^2 u = f \quad (6)$$

on the unit square $\Omega = [0,1] \times [0,1]$ with Dirichlet boundary conditions $u = 0$ on $\partial \Omega$.

An analytic solution used in [4], is the polynomial

$$u(x_1, x_2) = -x_1^2 x_2^2 (1 - x_1^2)(1 - x_2^2), \quad (7)$$

corresponding to the source

$$f(x_1, x_2) = -2(x_2^2(1 - 6x_1^2)(1 - x_2^2) + x_1^2(1 - 6x_2^2)(1 - x_1^2)). \quad (8)$$

The error of the approximations is computed using the normalized $L_1$-error

$$\frac{||u - v_h||_{L_1}}{||u||_{L_1}} = \frac{1}{\int_\Omega |u| d\Omega} \int_\Omega |u - v_h| d\Omega. \quad (9)$$

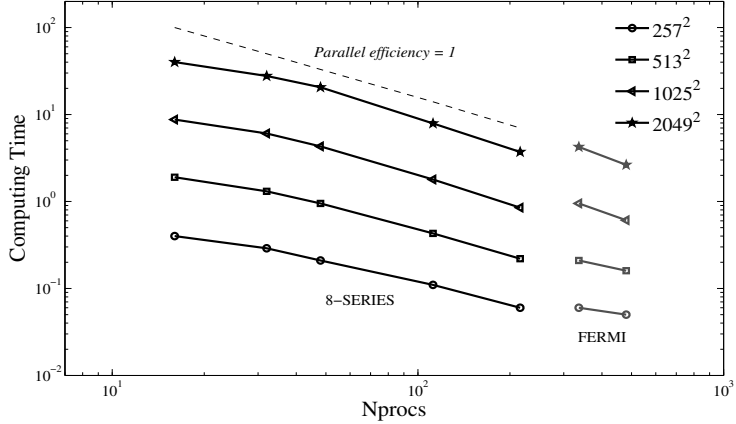

Figure 4: Computing time in seconds for a single cycle in GPUs with different architectures and number of processors (Nprocs). The implementation shows 100% parallel efficiency for large grids and large number of processors. The algorithm scales with every new generation of many-cores.

The recurrence of the cycles is shown in Figure 3 with maximum relaxations per level $n_r = 2$. The residual is computed using the maximum norm divided by the maximum norm of the sources. It shows a convergent saw pattern for every recurrence of the cycle, converging to machine precision. The $L_1$-error shows fast convergence for every cycle. Converging a little less than $10^{-2}$ for every cycle and stagnating to the discretization error. The diagnostic residual is the coarsened residual computed while applying the relaxation at every level with restricted sources and coefficients. This pseudo-residual is cost-less and is used to have control over the convergence of the algorithm in cases that the solution is unknown, by establishing a tolerance. Notice that for the full mesh, both residuals are the same. The examples shown in the following sections have reached a tolerance of $10^{-14}$.

The SGML algorithm is programmed with independent threads for every node in the grid, a straight forward formulation in many-core architectures for 100% parallel efficiency [15]. This has been observed in the numerical experiments shown in Figure 4. The runs are performed in GPUs with different number of processors.

## 3.2 Grid deformation

The deformation of grids is achieved using the *force potential* technique similar to that presented in [14]. A potential is obtained solving the Helmholtz equation

$$\nabla^2 u + au = f - \frac{1}{V} \int f d\Omega, \qquad (10)$$

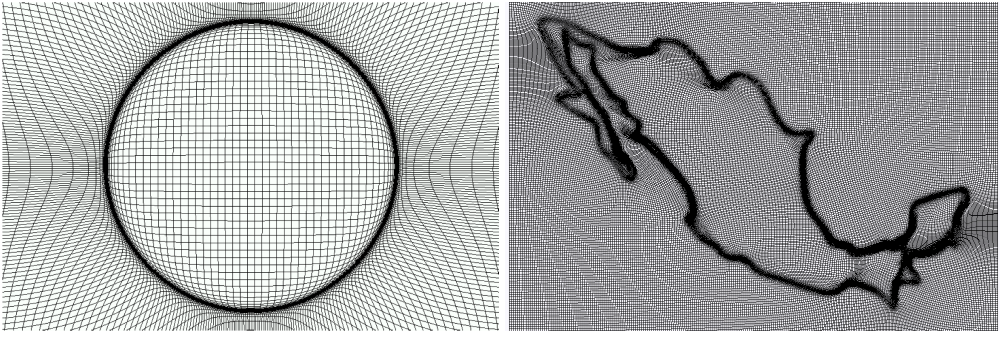

Figure 5: Grid deformation for a singular source with the shape of a circle in a $65^2$ mesh (left) and for the shape of Mexico in a $129^2$ mesh (right).

where $V$ is the area of $\Omega$, with Neumann boundary conditions $\frac{\partial u}{\partial n} = 0$.

The grid nodes are moved spatially with a forward Euler method, using equations of motion given by $\frac{d\mathbf{x}}{dt} = \mathbf{v}$, where the velocity

$$\mathbf{v} = \frac{-\nabla u}{tf + \int f d\Omega}, \qquad (11)$$

is computed using finite differences of the potential and with a pseudo-time $t$.

The source $f$ is a delta function over a parametrized curve, given by a set of points, not necessarily equidistant and resampled to equidistant points over the curve using the mesh spacing $h$. Z-splines [2] of first, third and fifth order are used to represent the continuous curve with the possibility of having discontinuous tangents. Tensor products of hat functions are used to represent a finite delta function source on the Cartesian grid. Examples for the deformation of a two-dimensional grid using the contours of a circle and Mexico are shown in Figure 5 with $a = 0.1$.

## 3.3 Streamlines of a singular trifoil-knot vortex

A three dimensional version of the SGML algorithm is used to solve the vector system

$$\nabla^2 \boldsymbol{\psi} = -\boldsymbol{\omega}, \qquad (12)$$

where $\boldsymbol{\psi} \in \mathbb{R}^3$ is the stream vector potential and $\boldsymbol{\omega}$ is the vorticity $\boldsymbol{\omega} = \nabla \times \mathbf{v}$, where $\mathbf{v}$ is the velocity field.

A singular one-dimensional source of vorticity, pointing tangentially, is placed in the form of a trifoil-knot described by

$$\begin{aligned} x_1(\theta) &= r(\sin\theta + 2\sin(2\theta)), \\ x_2(\theta) &= r(\cos\theta - 2\cos(2\theta)), \\ x_3(\theta) &= -2r\sin(3\theta), \end{aligned} \qquad (13)$$

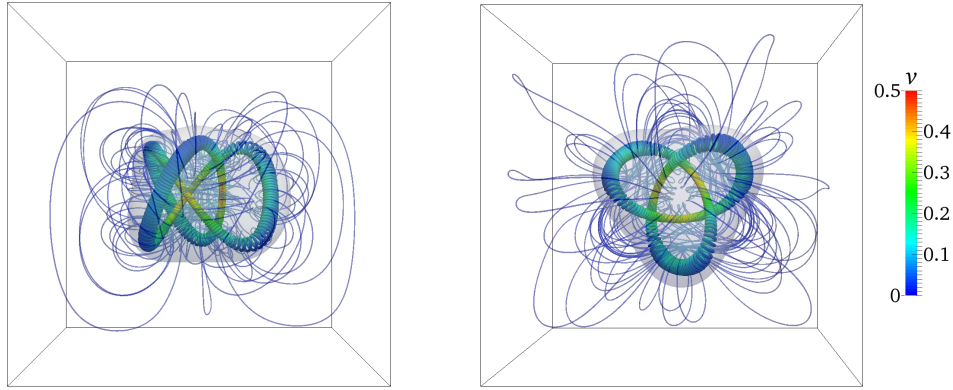

Figure 6: Lateral and frontal views of two streamlines around a singular trifoil-knot vortex colored with the scaled velocity magnitude $v$ for a mesh of $257^3$ nodes. One trajectory is trapped around the knot and the other orbits inside the box showing an average trifoil-orbit. A surface for $v = 0.02$ is shown in translucent grey.

where the parameter $\theta \in [0, 2\pi)$.

The velocity field can be obtained from the vector potential by

$$\boldsymbol{v} = \nabla \times \boldsymbol{\psi}. \tag{14}$$

Streamlines can be integrated using the resulting velocity field with a RK-marching scheme provided in the visualization tool.

The solutions for the vector potential and the streamlines of a trifoil-knot vortex in a $[0,1]^3$ box are shown in Figure 6 for a $257^3$ mesh and $r = 0.2$. Only two streamlines are shown: one close to the knot vortex, turning and traveling around the knot; and the second orbiting far from the singularity.

## 3.4 Heterogeneous coefficients

The heterogeneous Laplace equation

$$\nabla \cdot (\sigma \nabla u) = 0, \tag{15}$$

is solved in three dimensions for two smooth coefficients

$$\sigma(x_1, x_2, x_3) = 0.55 \pm 0.45 \tanh((r - 0.2)/0.1), \tag{16}$$

where $r$ is the distance from the centerpoint of a $[0,1]^3$ domain. Dirichlet conditions are imposed on the top and bottom boundaries, set to $u = 1$ and $u = -1$, respectively, and Neumann conditions $\partial u/\partial n = 0$ are used for all the lateral faces.

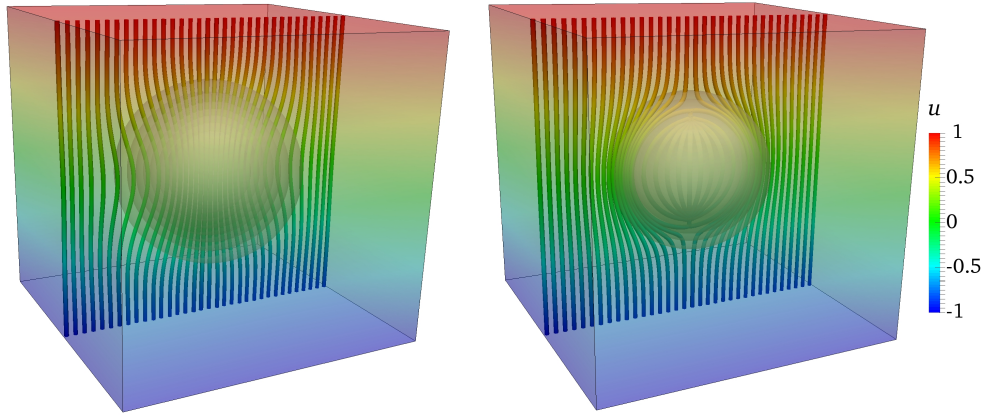

Figure 7: Force lines produced by the electric potential *u*, solution to the heterogeneous Laplace equation (15), for the spherical variable coefficient of conductivity $\sigma$ (16), with higher (left) and lower (right) conductivity than its surroundings. The resolution is $129^3$ and the conductivity is represented by transparent surfaces from 0.2 to 0.8.

The problem represents an electric potential *u* in a three dimensional capacitor filled with dielectric material of variable conductivity $\sigma$. Figure 7 shows the force lines $F = \nabla u$ produced by the electric potential *u* for the given conductivity (16), which is a spherical distribution with lower (+) and higher (-) conductivity than its surroundings. The pseudo-time step is adjusted locally according to the local coefficient using the stability condition $\delta\tau < (2^\nu h)^2/2\sigma$.

## 4   Conclusion

The SGML algorithm is a novel and scalable iterative solver, designed to run entirely inside the GPU without memory transfers to the CPU. It is simple to implement and provides an attractive option for the solution of elliptic partial differential equations in simulations running entirely inside GPUs, stand-alone or as a system of hybrid hyperbolic-elliptic equations. In particular, we are including the SGML algorithm in incompressible fluid simulations and in computations of flows in porous media with an implicit formulation for the pressure, requiring the solution of a Helmholtz-Poisson equation for every time step. The main difference with a classic multi-grid algorithm is the use of a single grid data structure combined with fine-grain parallelism. The routines operate over full-grid data, keeping a constant high occupancy of the streaming processors in the GPU. The algorithm has been proven to scale with every new generation of many-core architectures. The choose of the finite difference operator is important. We have observed that simple finite difference molecules could lead to divergence of the algorithm due to lack of correct information transmission for some directions of the mesh. Novel radial finite differences for the heterogeneous Helmholtz-Poisson equations have shown to provide convergence to machine precision for that particular equation. The overall residual is available from the algorithm data and can be used to monitor its convergence. Immersed interfaces could be used to obtain a solver for complex geometries without the need of special grids.

# Acknowledgements

This work was partially supported by ABACUS, CONACyT grant EDOMEX-2011-C01-165873.